\documentstyle[12pt]{article}
\begin{document}
\title{\vskip-3cm{\baselineskip14pt
\centerline{\normalsize DESY 00-068\hfill ISSN 0418-9833}
\centerline{\normalsize hep-ph/0004267\hfill}
\centerline{\normalsize April 2000\hfill}}
\vskip1.5cm
Order $\alpha^3\ln(1/\alpha)$ Corrections to Positronium Decays} 
\author{{\sc Bernd A. Kniehl} and {\sc Alexander A. Penin}\thanks{Permanent
address: Institute for Nuclear Research, Russian Academy of Sciences,
60th October Anniversary Prospect 7a, Moscow 117312, Russia.}\\
{\normalsize II. Institut f\"ur Theoretische Physik, Universit\"at Hamburg,}\\
{\normalsize Luruper Chaussee 149, 22761 Hamburg, Germany}}

\date{}

\maketitle

\thispagestyle{empty}

\begin{abstract}
The logarithmically enhanced $\alpha^3\ln(1/\alpha)$ corrections to the
para- and orthopositronium decay widths are calculated in the framework of
dimensionally regularized nonrelativistic quantum electrodynamics. 
In the case of parapositronium, the correction is negative, approximately 
doubles the effect of the leading logarithmic $\alpha^3\ln^2(1/\alpha)$ one,
and is comparable to the nonlogarithmic $O(\alpha^2)$ one.
As for orthopositronium, the correction is positive and almost cancels the
$\alpha^3\ln^2(1/\alpha)$ one.
The uncertainties in the theoretical predictions for the decay widths are
reduced.
\medskip

\end{abstract}

\newpage

Positronium (Ps), which is an electromagnetic bound state of the electron
$e^-$ and the positron $e^+$, is the lightest known atom.
Since its theoretical description is not plagued by strong-interaction
uncertainties, thanks to the smallness of the electron mass $m_e$ relative to
typical hadronic mass scales, its properties can be calculated perturbatively
in quantum electrodynamics (QED), as an expansion in Sommerfeld's
fine-structure constant $\alpha$, with very high precision.
Ps is thus a unique laboratory for testing the QED theory of weakly bound
systems.

The decay widths of the $^1S_0$ parapositronium (p-Ps) and $^3S_1$
orthopositronium (o-Ps) ground states to two and three photons, respectively,
have been the subject of a vast number of theoretical and experimental
investigations.  
The present theoretical knowledge may be summarized as:
\begin{eqnarray}
\Gamma_p^{\rm th}&=&\Gamma^{(0)}_{p}
\left[1+A_p{\alpha\over\pi}+2\alpha^2\ln{1\over\alpha}
+B_p\left({\alpha\over\pi}\right)^2
\right.\nonumber\\
&&{}-\left.
{3\alpha^3\over2\pi}\ln^2{1\over\alpha} 
+C_p{\alpha^3\over\pi}\ln{1\over\alpha}+D_p\left({\alpha\over\pi}\right)^3
+\ldots\right],
\label{spar}\\
\Gamma_o^{\rm th}&=&\Gamma^{(0)}_{o}
\left[1+A_o{\alpha\over\pi}-{\alpha^2\over3}\ln{1\over\alpha}
+B_o\left({\alpha\over\pi}\right)^2 
\right.\nonumber\\
&&{}-\left.
{3\alpha^3\over2\pi}\ln^2{1\over\alpha}
+C_o{\alpha^3\over\pi}\ln{1\over\alpha}+D_o\left({\alpha\over\pi}\right)^3
+\ldots\right],
\label{sort}
\end{eqnarray}
where 
\begin{eqnarray}
\Gamma^{(0)}_{p}&=&{\alpha^5m_e\over 2},\nonumber\\
\Gamma^{(0)}_{o}&=&{2(\pi^2-9)\alpha^6m_e\over 9\pi},
\end{eqnarray}
are the lowest-order results.
The $O(\alpha)$ coefficients in Eqs.~(\ref{spar}) \cite{HarBra} and
(\ref{sort}) \cite{CLS} read 
\begin{eqnarray}
A_p&=&{\pi^2\over4}-5,\nonumber\\
A_o&=&-10.286\,606(10).
\end{eqnarray}
The logarithmically enhanced $\alpha^2\ln(1/\alpha)$ terms in 
Eqs.~(\ref{spar}) and (\ref{sort}) have been obtained in
Refs.~\cite{KhrYel,CasLep1}, respectively.
Recently, the nonlogarithmic $O(\alpha^2)$ coefficients in Eqs.~(\ref{spar}) 
\cite{CMY2} and (\ref{sort}) \cite{AFS} have been found to be
\begin{eqnarray}
B_p&=&5.14(30),\nonumber\\
B_o&=&44.52(26).
\label{bpbo}
\end{eqnarray}
Note that the light-by-light scattering diagrams have been 
omitted in Ref.~\cite{AFS}. In the p-Ps calculation \cite{CMY2}
the diagrams of this type increase the coefficient
$B_p$ by $1.28$, and, therefore, their contribution to 
the coefficient $B_o$ is assumed to be relatively small.
The p-Ps (o-Ps) decays into four (five) photons, which are not included in 
Eq.~(\ref{bpbo}), lead to an increase of the coefficient $B_p$ ($B_o$) by
$0.274(1)$ ($0.19(1)$) \cite{LMSZ}.
In $O(\alpha^3)$, only the leading logarithmic $\alpha^3\ln^2(1/\alpha)$ terms
are known \cite{Kar}.
Including all the terms known so far, we obtain for the p-Ps and o-Ps total
decay widths
\begin{eqnarray}
\Gamma_p^{\rm th}&=&7989.512(13)~\mu{\rm s}^{-1},
\label{thpar}\\
\Gamma_o^{\rm th}&=&7.039943(10)~\mu{\rm s}^{-1},
\label{thort}
\end{eqnarray}
where the given errors stem only from the coefficients $B_p$ 
and $B_o$ respectively and we postpone  the discussion of total
uncertainty of theoretical estimates to the end of the paper.
The purpose of this letter is complete our knowledge of the logarithmically 
enhanced terms of $O(\alpha^3)$ by providing the coefficients $C_p$ and $C_o$,
in analytic form.
We also give order-of-magnitude estimates of the unknown coefficients $D_p$
and $D_o$.

On the experimental side, the present situation is not entirely clear.
Recently, the Ann Arbor group measured the p-Ps width to be \cite{Annp}
\begin{equation}
\Gamma_p^{\rm exp}=7990.9(1.7)~\mu{\rm s}^{-1},
\label{exannp}
\end{equation} 
which agrees with Eq.~(\ref{thpar}) within the experimental error.
However, in the case of o-Ps, their measurements \cite{Ann1o,Ann2o},
\begin{eqnarray}
\Gamma_o^{\rm exp}(\mbox{gas})&=&7.0514(14)~\mu{\rm s}^{-1},  
\nonumber\\
\Gamma_o^{\rm exp}(\mbox{vacuum})&=&7.0482(16)~\mu{\rm s}^{-1},  
\label{exanno} 
\end{eqnarray}
exceed Eq.~(\ref{thort}) by 8 and 5 experimental standard deviations,
respectively.
This apparent contradiction is known as the {\it o-Ps lifetime puzzle}.
On the other hand, the Tokyo group found \cite{Tok}  
\begin{eqnarray}
\Gamma_o^{\rm exp}(\mbox{SiO$_2$})&=&7.0398(29)~\mu{\rm s}^{-1},
\label{extoko}
\end{eqnarray}
which agrees with Eq.~(\ref{thort}) within the experimental error.
Leaving this aside, the o-Ps results from Ann Arbor could be considered as a
signal of new physics beyond the standard model.
However, a large number of exotic decay modes have already been ruled out
\cite{CzaKar}.
No conclusion on the o-Ps lifetime puzzle can be drawn until the experimental
precision increases and the data become unambiguous.

On the theoretical side, it is an urgent matter to improve the predictions of
the Ps lifetimes as much as possible.
Thus, one is faced with the task of analyzing the $O(\alpha^3)$ corrections,
which is extremely difficult, especially for o-Ps.
However, there is a special subclass of the $O(\alpha^3)$ corrections which
can be analyzed separately, namely those which are enhanced by powers of
$\ln(1/\alpha)\approx5$.
They may reasonably be expected to provide an essential
part  of the $O(\alpha^3)$  corrections.
This may be substantiated by considering Eqs.~(\ref{spar}) and (\ref{sort}) in
$O(\alpha^2)$, where logarithmic terms enter for the first time.
In the case of p-Ps, 98\% of the $O(\alpha^2)$ correction stems from the 
logarithmic term. In the case of o-Ps the logarithmic term is not
so dominant but still gives about $1/4$  of the total $O(\alpha^2)$
correction.  
The origin of the logarithmic corrections is the presence of several scales
in the bound-state problem.
The dynamics of the nonrelativistic (NR) $e^+e^-$ pair near threshold involves
four different scales \cite{BenSmi}:
(i) the hard scale (energy and momentum scale like $m_e$);
(ii) the soft scale (energy and momentum scale like $\beta m_e$);
(iii) the potential scale (energy scales like $\beta^2m_e$, while 
momentum scales like $\beta m_e$); and
(iv) the ultrasoft (US) scale (energy and momentum scale like $\beta^2m_e$).
Here $\beta$ denotes the electron velocity in the center-of-mass frame.
The logarithmic integration over a loop momentum between different scales 
yields a power of $\ln(1/\beta)$.
Since Ps is approximately a Coulomb system, we have $\beta\propto\alpha$. 
This explains the appearance of powers of $\ln(1/\alpha)$ in 
Eqs.~(\ref{spar}) and (\ref{sort}).
The leading logarithmic corrections may be obtained straightforwardly by
identifying the regions of logarithmic integration \cite{KhrYel,CasLep1,Kar}.
The calculation of the subleading logarithms is much more involved because
certain loop integrations must be performed exactly beyond the logarithmic
accuracy.

In the following, we briefly outline the main features of our analysis.
We work in NR QED (NRQED) \cite{CasLep2}, which is the effective field theory
that emerges by expanding the QED Lagrangian in $\beta$ and integrating out
the hard modes.
If we also integrate out the soft modes and the potential photons, we arrive
at the effective theory of potential NRQED (pNRQED) \cite{PinSot1}, which
contains potential electrons and US photons as active particles.
Thus, the dynamics of the NR $e^+e^-$ pair is governed by the effective
Schr\"odinger equation and by its multipole interaction with the US photons.
The corrections from harder scales are contained in the higher-dimensional
operators of the NR Hamiltonian, corresponding to an expansion in $\beta$, and
in the Wilson coefficients, which are expanded in $\alpha$. 
In the process of scale separation, spurious infrared (IR) and ultraviolet
(UV) divergences arise, which endow the operators in the NR Hamiltonian with
anomalous dimensions.
We use dimensional regularization (DR), with $d=4-2\epsilon$ space-time 
dimensions, to handle these divergences \cite{PinSot1,PinSot2,CMY1}.
This has the advantage that contributions from different scales are matched
automatically.
The logarithmic corrections are closely related to the anomalous dimensions
and can be found by analyzing the divergences of the NR effective theory.
In this way, we have obtained the leading logarithmic third-order corrections
to the energy levels and wave functions at the origin of heavy quark-antiquark
bound states \cite{KniPen2}, which includes the QED result
\cite{KhrYel,CasLep1,Kar} as a special case.
Here, we extend this approach to the subleading logarithms in QED.
Note that the NRQED approach, endowed with an explicit momentum cutoff and a
fictitious photon mass to regulate the UV and IR divergences, has also been
applied to find the third-order correction, including subleading logarithms,
to the hyperfine splitting in muonium \cite{NioKin}.

The annihilation of Ps is the hard process which gives rise to imaginary parts
in the local operators of the NR Hamiltonian \cite{LLM}.
The decay width can be obtained by averaging these operators over the
bound-state wave function.
The hard-scale corrections, which require fully relativistic QED calculations
and are most difficult to find, do not depend on $\beta$ and do not lead to
logarithmic contributions by themselves.
However, they can interfere with the logarithmic corrections from the softer
scales.
Thus, the only results from relativistic perturbation theory that enter our
analysis are (i) the one-loop hard renormalizations of the imaginary parts of
the leading four-fermion operators, {\it i.e.}, the Born decay amplitudes,
which are given by the coefficients $A_p$ and $A_o$, and (ii) the hard parts
of the one-loop $O(\alpha\beta^2)$ operators \cite{MPS}.
The missing ingredients can all be obtained in the NR approximation.
These include (i) the correction to the Ps ground-state wave function at the
origin due to the $O(\alpha\beta^2)$ terms in the NR Hamiltonian, (ii) the
$O(\alpha\beta^2)$ and $O(\alpha^2\beta)$ corrections to the leading
four-fermion operators, and (iii) the correction due to the emission and
absorption of US photons by the Ps bound state.

The value of the ground-state ($n=1$) wave function at the origin $\psi_1(0)$
may be extracted from the NR Green function $G({\bf x},{\bf y},E)$, which
satisfies the equation
\begin{equation}
\left({\cal H}_C+\Delta{\cal H}-E\right)G({\bf x},{\bf y},E)
=\delta^{(3)}({\bf x}-{\bf y}),
\label{Sch}
\end{equation}
where ${\cal H}_C$ is the Coulomb Hamiltonian and $\Delta{\cal H}$ stands for
the terms of higher orders in $\alpha$ and $\beta$.
The solution of Eq.~(\ref{Sch}) can be found in time-independent perturbation 
theory as an expansion in $\alpha$ around the leading-order Coulomb Green
function.
We thus obtain the correction $\Delta\psi_1^2$ in the relationship
$|\psi_1(0)|^2=\left|\psi_1^C(0)\right|^2\left(1+\Delta\psi_1^2\right)$, where
$\psi_1^C(0)$ is the ground-state wave function at the origin in the Coulomb
approximation.
As mentioned above, this analysis may be enormously simplified by the use of
DR.
Proceeding along the lines of Ref.~\cite{KniPen2}, we thus recover with ease
the well-known $\alpha^2\ln(1/\alpha)$ terms in Eqs.~(\ref{spar}) and
(\ref{sort}) \cite{KhrYel,CasLep1},
\begin{equation}
\Delta^\prime\psi_1^2={\alpha^2}\ln{1\over\alpha}
\left[2-{7\over 6}S(S+1)\right],
\label{wfcorp}
\end{equation}
where $S$ is the eigenvalue of the total-spin operator ${\bf S}$.
As mentioned above, $\Delta^\prime\psi_1^2$ interferes with the one-loop hard
renormalization of the Born amplitudes, $A_p$ and $A_o$, to produce
$\alpha^3\ln(1/\alpha)$ terms.
The resulting contributions to the coefficients $C_p$ and $C_o$ read $2A_p$
and $-A_o/3$, respectively.

The generic logarithmic $O(\alpha^3)$ correction
$\Delta^{\prime\prime}\psi_1^2$ to $|\psi_1(0)|^2$ is generated by the
following one-loop operators, given in the momentum representation with
$O(\epsilon)$ accuracy,
\begin{eqnarray}
\Delta_{\rm h}{\cal H}&=&-{1\over3}\,{\alpha^2\over m_e^2}
\left[{1\over\hat\epsilon}\left({\mu^2\over m_e^2}\right)^{\epsilon}
+{39\over5}-12\ln2
\right.\nonumber\\
&&{}+\left.
\left({32\over3}+6\ln2\right){\bf S}^2\right],
\label{hamhard}\\
\Delta_{\rm s}{\cal H}&=&-{7\over3}\,{\alpha^2\over m_e^2}
\left[{1\over\hat\epsilon}\left({\mu^2\over{\bf q}^2}\right)^{\epsilon}
-{1\over7}\right],
\label{hamsoft}\\
\Delta_{\rm us}{\cal H}&=&{8\over3}\,{\alpha^2\over m_e^2}
\left[{1\over\hat\epsilon}\left({\mu\over{\bf p}^2/m_e-E_1^C}
\right)^{2\epsilon}+{5\over3}-2\ln2\right],
\label{hamus}
\end{eqnarray}
where $1/\hat\epsilon=1/\epsilon-\gamma_E+\ln(4\pi)$, with $\gamma_E$ being
Euler's constant, $\mu$ is the 't~Hooft mass scale of DR, ${\bf q}$ is the
three-momentum transfer, and $E_1^C=-\alpha^2m_e/4$ is the Coulomb
ground-state energy.
Equations~(\ref{hamhard}) and (\ref{hamsoft}) give the hard \cite{MPS} and
soft \cite{PinSot2} $O(\alpha\beta^2)$ contributions to the NR Hamiltonian,
respectively.
The US contribution given in Eq.~(\ref{hamus}) arises from the emission and
absorption of an US photon, which converts the on-shell Ps ground state into
some off-shell state of the Coulomb spectrum, with three-momentum ${\bf p}$,
before it decays.
It is the only US contribution which can be represented by an operator of
instantaneous interaction and thus give rise to logarithmic corrections.
It has been found with the help of the method developed for the more
complicated case of quantum chromodynamics in Ref.~\cite{KniPen1}, where it 
was applied to the on-shell renormalization of the heavy-quarkonium wave
function at the origin.
The singularities of the operators in Eqs.~(\ref{hamhard})--(\ref{hamus})
yield the logarithmic corrections which we are interested in.
Up to their logarithmic dependences on ${\bf q}^2$ and ${\bf p}^2$, these
operators are of the $\delta$-function type in coordinate space and,
therefore, lead to additional singularities in the Coulomb Green function at
the origin \cite{KniPen2}.
As a consequence, in the evaluation of the Green function in time-independent
perturbation theory from Eq.~(\ref{Sch}) with $\Delta_{\rm h}{\cal H}$,
$\Delta_{\rm s}{\cal H}$, and $\Delta_{\rm us}{\cal H}$, overlapping 
logarithmic divergences appear in the part of the first-order term which
corresponds to the interference of the one-photon contribution to the Coulomb
Green function and the first terms of Eqs.~(\ref{hamhard})--(\ref{hamus}). 
This results in the double-logarithmic contribution, which can be directly
extracted from the coefficient of the leading double-pole singularity
\cite{KniPen2}.  
Since we are interested in the single-logarithmic contribution, we also have 
to keep the subleading terms in this analysis.
The logarithmic corrections which are generated by the non-overlapping 
singularities can be obtained by putting $\mu=m_e$ in the Coulomb Green 
function at the origin and ${\bf q}^2={\bf p}^2=-m_eE_1^C$ in
Eqs.~(\ref{hamhard})--(\ref{hamus}), and proceeding as in the evaluation of
Eq.~(\ref{wfcorp}).
We thus obtain
\begin{eqnarray}
\Delta^{\prime\prime}\psi_1^2&=&{\alpha^3\over\pi}
\left\{-{3\over2}\ln^2{1\over\alpha}+\left[-{184\over45}+{2\over3}\ln2
\right.\right.\nonumber\\
&&{}+\left.\left.
\left({16\over9}+\ln2\right)S(S+1)\right]\ln{1\over\alpha}\right\}.
\end{eqnarray}
The first term herein agrees with the corresponding terms in Eqs.~(\ref{spar})
and (\ref{sort}) \cite{Kar}, while the second one represents a new result.

Another source of $\alpha^3\ln(1/\alpha)$ terms is the $O(\alpha\beta^2)$ 
corrections to the leading four-fermion operators.
Since they do not involve the singular Coulomb Green function at the origin,
there are no overlapping divergences, and we may simply read off the resulting
$\alpha^3\ln(1/\alpha)$ terms from the poles of their US parts, which are
given by the operator 
\begin{equation}
-{1\over\epsilon}\,{2\alpha\over3\pi}\,
{{\bf p}^2+{\bf p^\prime}^2\over m_e^2}V_4({\bf p},{\bf p^\prime},{\bf S}).
\label{power}
\end{equation}
Here $V_4({\bf p},{\bf p^\prime},{\bf S})$ is the local four-fermion operator
which generates the leading-order decay widths.
Taking the expectation value of Eq.~(\ref{power}) w.r.t.\ the ground-state
wave function, one encounters power-divergent integrals \cite{LLM}.
They can be consistently treated within DR \cite{CMY2}.
This leads to the substitution ${\bf p}^2,{\bf p^\prime}^2\to m_eE_1^C$ in the
matrix element.
The UV-pole contribution of Eq.~(\ref{power}) is then canceled by the IR pole
of the hard contribution \cite{Kur}.
This implies that the logarithmic integration ranges from the US scale
$\alpha^2m_e$ up to the hard scale $m_e$, so that $1/\epsilon$ should be
replaced by $4\ln(1/\alpha)$ \cite{KniPen2}.
The resulting $\alpha^3\ln(1/\alpha)$ corrections to the decay widths are
spin-independent and read
\begin{equation}
\Delta\Gamma_{p,o}=\Gamma^{(0)}_{p,o}{4\alpha^3\over3\pi}\ln{1\over\alpha}.
\end{equation}

The last source of $\alpha^3\ln(1/\alpha)$ terms is the $O(\alpha^2\beta)$ 
corrections to the leading four-fermion operators.
These corrections are non-analytic in ${\bf p}^2$ and of the form 
\begin{equation}
\left[-{7\over 6}+\left(8-{32\over3}\ln2\right)\right]{\alpha^2\over4}\,
{|{\bf p}|+|{\bf p^\prime}|\over m_e}V_4({\bf p},{\bf p^\prime},{\bf S}),
\label{powernonan}
\end{equation}
where the first and second terms contained within the square brackets are the
soft and US contributions, respectively.
Although the coefficient of $V_4({\bf p},{\bf p^\prime},{\bf S})$ in
Eq.~(\ref{powernonan}) is finite, the matrix element of Eq.~(\ref{powernonan})
between Ps bound-state wave functions is logarithmically divergent in the UV
region.
In DR, the divergent part is given by the matrix element of 
Eq.~(\ref{powernonan}) with $|{\bf p}|,|{\bf p^\prime}|$ replaced by
$m_e\alpha/(\pi\epsilon)$.  
In contrast to Eq.~(\ref{power}), the logarithmic integration now ranges from
the soft scale up to the hard one, and $1/\epsilon$ should be replaced by
$2\ln(1/\alpha)$. 
The resulting  contributions to the decay widths read
\begin{equation}
\Delta\Gamma_{p,o}=\Gamma^{(0)}_{p,o}{\alpha^3\over\pi}\ln{1\over\alpha}
\left({41\over 6}-{32\over 3}\ln2\right).
\end{equation}

Summing up the various $\alpha^3\ln(1/\alpha)$ terms derived above, we obtain 
\begin{eqnarray}
C_p&=&2A_p+{367\over90}-10\ln2\approx-7.919,\nonumber\\
C_o&=&-{A_o\over 3}+{229\over 30}-8\ln2\approx5.517.
\end{eqnarray}
In the case of p-Ps, the new $\alpha^3\ln(1/\alpha)$ term has the same sign 
as the $\alpha^3\ln^2(1/\alpha)$ one and exceeds the latter in magnitude.
The sum of these two terms compensates approximately 1/3 of the positive
contribution from the nonlogarithmic $O(\alpha^2)$ term.
As for o-Ps, the new $\alpha^3\ln(1/\alpha)$ term cancels approximately 3/4
of the $\alpha^3\ln^2(1/\alpha)$ contribution.
Our final predictions for the p-Ps and o-Ps total decay widths, including the 
multi-photon channels, read
\begin{eqnarray}
\Gamma_p^{\rm th}&=&7989.620(13)~\mu{\rm s}^{-1},
\label{finpar}\\
\Gamma_o^{\rm th}&=&7.039968(10)~\mu{\rm s}^{-1},
\label{finort}
\end{eqnarray}
which has to be compared with Eqs.~(\ref{thpar}) and (\ref{thort}).
As before, Eq.~(\ref{finpar}) agrees with the Ann Arbor \cite{Annp}
measurement~(\ref{exannp}), and Eq.~(\ref{finort}) favours the Tokyo
\cite{Tok} measurement~(\ref{extoko}), while it significantly undershoots the
Ann Arbor \cite{Ann1o,Ann2o} measurements~(\ref{exanno}).

The missing nonlogarithmic $O(\alpha^3)$ corrections in Eqs.~(\ref{spar}) and
(\ref{sort}) receive contributions from three-loop QED diagrams with a
considerable number of external lines, which are far beyond the reach of
presently available computational techniques.
In this sense, we expect Eqs.~(\ref{finpar}) and (\ref{finort}) to remain the
best predictions for the forseeable future.
However, we may speculate about the magnitudes of the coefficients $D_p$ and
$D_o$.
Two powers of $\alpha$ in these terms can be of NR origin.
Each of them should be accompanied by the characteristic factor $\pi$, which
happens for the logarithmic terms.
Thus, we estimate the coefficients $D_p$ and $D_o$ to be a few units times
$\pi^2$.
This rule of thumb is in reasonable agreement with the situation at
$O(\alpha^2)$, where we have $B_p\approx\pi^2/2$ and $B_o\approx4\pi^2$.
If the coefficients $D_p$ and $D_o$ do not have magnitudes in excess of 100,
then the uncertainties due the lack of their knowledge falls within the errors
quoted in Eqs.~(\ref{finpar}) and (\ref{finort}).
Then, our new results reduce the uncertainties in the predicted p-Ps 
decay width to $10^{-2}\mu{\rm s}^{-1}$.
The main remaining theoretical uncertainty 
in o-Ps decay width is related to 
the unknown $O(\alpha^2)$ contribution of 
the light-by-light scattering diagrams
which can be estimated as  a few units times $10^{-5}\mu{\rm s}^{-1}$
on the basis of p-Ps result. Calculation of this contribution
along with our present result will reduce   
the uncertainties in the predicted o-Ps 
decay width to $10^{-5}\mu{\rm s}^{-1}$.
Further progress in our understanding the Ps lifetime problem crucially
depends also on the reduction of the experimental errors, 
which now greatly exceed the theoretical ones.

Finally, we note that the technique developed in this letter can also be
applied to the calculation of the subleading logarithmic
$\alpha^7\ln(1/\alpha)$ terms for the Ps hyperfine splitting.
This problem is of special interest because of the apparent discrepancy
between the latest experimental data \cite{HFS} and the best theoretical
predictions, which include the $O(\alpha^6)$ corrections (see Ref.~\cite{CMY1}
and references cited therein) and the leading logarithmic
$\alpha^7\ln^2\alpha$ term \cite{Kar}.

We are indebted to R. Hill and G. P. Lepage for useful discussions and to
K. Melnikov and A. Yelkhovsky for pointing out the relevance of the
$O(\alpha^2\beta)$ contribution from Eq.~(\ref{powernonan}).
This work was supported in part by the Deutsche Forschungsgemeinschaft under
Contract No.\ KN~365/1-1, by the Bundesministerium f\"ur Bildung und Forschung
under Contract No.\ 05~HT9GUA~3, and by the European Commission through the
Research Training Network {\it Quantum Chromodynamics and the Deep Structure
of Elementary Particles} under Contract No.\ ERBFMRXCT980194.
The work of A.A.P. was supported in part by the Volkswagen Foundation under
Contract No.\ I/73611.

\smallskip

{\it Note added}

After including the $O(\alpha^2\beta)$ contribution from
Eq.~(\ref{powernonan}), which was missed in the previous version of this
letter, we find agreement with the analytical results for the p-Ps and o-Ps
decay widths of Ref.~\cite{MY} and with the numerical result for the o-Ps
decay width of Ref.~\cite{HL}.
However, this correction is of insignificant size and immaterial for our
numerical estimates of the Ps decay widths.

\end{document}